\documentclass[12pt,a4paper]{article}

\usepackage[T1]{fontenc}
\usepackage{lmodern}
\usepackage{amsmath,amssymb,amsthm,mathtools}
\usepackage{geometry}
\usepackage{url}
\usepackage{microtype}
\usepackage{graphicx}
\usepackage{booktabs}
\usepackage[hidelinks,bookmarks=false]{hyperref}

\numberwithin{equation}{section}

\newtheorem{thm}{Theorem}[section]
\newtheorem{prop}[thm]{Proposition}
\newtheorem{lem}[thm]{Lemma}
\newtheorem{cor}[thm]{Corollary}
\theoremstyle{remark}
\newtheorem{rem}[thm]{Remark}
\theoremstyle{definition}

\newcommand{\R}{\mathbb R}
\newcommand{\N}{\mathbb N}
\newcommand{\C}{\mathbb C}

\newcommand{\spec}{\sigma}
\newcommand{\ess}{\mathrm{ess}}
\newcommand{\Mfun}{M}
\newcommand{\Ufun}{U}

\title{Landau-level splitting by a circular $\delta$ interaction: boundary overlap and the hard-wall crossover}
\author{Masahiro Kaminaga\\
Tohoku Gakuin University, Sendai, Japan}
\date{}

\begin{document}
\maketitle

\begin{abstract}
A thin circular barrier lifts the angular momentum degeneracy of Landau
states according to their overlap with the barrier. We quantify this effect
for the two-dimensional Landau Hamiltonian with a circular $\delta$
interaction of radius $a$ and strength $\alpha$. Starting from the known
Kummer-function matching equation, we separate the contribution of one
Landau level from the remaining radial spectrum. For fixed $n$, $B>0$,
$a>0$, and $\alpha\ne0$, we prove
$$
E_{n,m}^{(\alpha)}-B(2n+1)
=\alpha c_{n,m}\left[1-\frac{\alpha a}{2m}+O(m^{-2})\right]
$$
as $m\to+\infty$, where $c_{n,m}$ is an explicit boundary trace
weight with factorial decay. The proof gives a uniform estimate of the
regular part of the radial resolvent, including at the Landau energy.
This estimate also yields an expansion uniform over all repulsive strengths,
with crossover parameter $\alpha a/(2m)$, and the exterior Dirichlet limit.
Thus a small energy shift does not by itself guarantee relative accuracy
of the projected energy shift: the correction depends on the wall strength
compared with the local centrifugal scale. The state nevertheless remains
close in norm to the chosen Landau subspace, uniformly for repulsion.
We also obtain the boundary density from the response to $\alpha$ and
distinguish nodal Landau states from coupling-dependent level crossings.
High-precision calculations test both the fixed-strength expansion and
the crossover to a hard wall.
\end{abstract}


\section{Introduction}\label{sec:intro}

A narrow annular barrier provides a simple setting in which to study how
Landau states respond to a localized boundary perturbation.  States whose
probability density is concentrated near the barrier can be shifted
appreciably, whereas states centered far from it interact with the barrier
only through a small tail of the wavefunction.  This raises two related
questions.  First, to what extent can the energy shift of an individual
Landau mode be predicted from its probability weight at the barrier?
Second, how does a finite penetrable barrier approach the hard-wall limit?
These questions involve not only the absolute size of the energy shift but
also the mixing with radial levels outside the chosen Landau subspace.

We consider a single spinless particle, without disorder or particle
interactions, on the full plane.  We use units
$\hbar=2M=|q|=1$, where $M$ and $q$ denote the particle mass and charge
magnitude, and choose the charge convention in $(-i\nabla-A)^2$.  Let
$B>0$ be the field parameter and let
$$
A(x)=\frac{B}{2}(-x_2,x_1)
$$
be the symmetric-gauge vector potential.  The free Landau Hamiltonian
$$
H_{B,0}=(-i\nabla-A)^2
$$
in $L^2(\R^2)$ has the spectrum
$$
\spec(H_{B,0})=\{\Lambda_n:n\in\N_0\},
\qquad
\Lambda_n=B(2n+1),
$$
and every $\Lambda_n$ is an eigenvalue of infinite multiplicity
\cite{LandauLifshitz}.

The spectral response of Landau levels to localized perturbations has been
studied in several different regimes.  For rapidly decreasing electric
potentials, Raikov and Warzel obtained nonclassical accumulation laws
\cite{RaikovWarzel2002}.  Melgaard and Rozenblum studied eigenvalue
asymptotics for constant magnetic fields in arbitrary even dimension
\cite{MelgaardRozenblum2003}, and Pushnitski, Raikov, and Villegas-Blas
later analyzed the asymptotic density of Landau clusters in the
high-Landau-level limit \cite{PushnitskiRaikovVillegas2013}.  For hard
obstacles, Pushnitski and Rozenblum obtained capacity-dependent
accumulation asymptotics for the exterior Dirichlet Landau problem
\cite{PushnitskiRozenblum2007}.  These results describe different aspects
of the way in which the infinite Landau degeneracy is resolved by
localized perturbations or boundaries.

In the present paper the perturbation is concentrated on a circle and is
modeled by a delta interaction.  Such curve-supported interactions give an
idealized description of a thin penetrable barrier.  The general
operator-theoretic framework for delta interactions supported on smooth
curves in a magnetic field is well established.  In particular, Behrndt,
Exner, Holzmann, and Lotoreichik \cite{BEHL2020} proved self-adjointness,
preservation of the essential Landau spectrum, one-sided eigenvalue
accumulation for sign-definite interactions, and capacity-dependent
accumulation estimates, and also established approximation by thin regular
potentials.  Behrndt, Holzmann, Lotoreichik, and Raikov
\cite{BHLR2022} subsequently studied eigenfunctions at the Landau energies
and the role of Laguerre zeros for circular support.

For a circular interaction, separation into angular momentum channels
reduces the problem to a radial matching equation.  A corresponding
Kummer-function formulation for magnetic soft rings was derived by Exner
and Tater \cite{ExnerTater2004}.  We use this established scalar matching
problem as a starting point.  Our concern here is different: we study the
mode-resolved energy shift at fixed Landau index, determine the correction
to the one-Landau-level projection, and follow the finite-strength problem
uniformly into the hard-wall regime.

More precisely, the coefficient of $(\Lambda_n-E)^{-1}$ in the radial
Green function is the boundary trace weight $c_{n,m}$ of the corresponding free
Landau state.  The contribution of the remaining radial spectrum is much
larger on an algebraic scale and has the expansion
$$
\frac{a}{2m}+\frac{ax}{2m^2}+O(m^{-3}),
\qquad
x=\frac{Ba^2}{2}.
$$
Combining the pole term with this regular contribution gives the relative
correction to the projected Landau energy for fixed interaction strength.
Retaining the regular contribution in the denominator also produces an
estimate uniform over all repulsive strengths and identifies the crossover
parameter
$$
\beta=\frac{\alpha a}{2m}.
$$
Thus a small absolute energy shift does not necessarily imply that
the projected energy shift is accurate in relative terms. The normalized
state can still remain close to the original Landau subspace; we quantify
this distinction in Proposition~\ref{prop:landau-weight}. The same
calculation also describes the saturation of the shift as the penetrable
barrier approaches an impenetrable boundary.

The physical origin of the large-$m$ regime is particularly transparent in
the symmetric gauge.  At fixed Landau index, increasing positive angular
momentum moves the radial probability distribution outward while leaving
the free cyclotron energy unchanged.  A wall of fixed radius therefore
samples only the tail of sufficiently large-$m$ states.  This produces a
factorially small boundary overlap and hence a factorially small absolute
shift.  At the same time, virtual coupling to the remaining radial levels
contributes on the algebraic scale $m^{-1}$.  The competition between these
two scales is what controls the relative projection error and the
penetrable-to-hard-wall crossover.

Our purpose is therefore not to derive the general spectral theory of
curve-supported interactions, but to resolve the Landau-level splitting by
angular momentum and to quantify the finite-strength correction for
individual modes.  All large-$m$ limits below keep $n$, $B$, and $a$ fixed.
They are consequently different from the high-Landau-level limit studied
in \cite{PushnitskiRaikovVillegas2013}.

An important point is the angular-momentum indexing.  In the symmetric
gauge, the free radial spectrum in channel $m\in\mathbb Z$ is
$$
B\bigl(2j+|m|-m+1\bigr),\qquad j\in\N_0.
$$
Hence a fixed Landau level $\Lambda_n$ occurs in the channel $m$ if and
only if $m\ge -n$.  In particular, the infinite degeneracy relevant to the
large-mode cluster is generated by $m\to+\infty$, not by
$|m|\to\infty$ symmetrically.  This distinction is important in the
asymptotic analysis.

The paper is organized as follows.  Section~\ref{sec:model} introduces the
operator by a quadratic form and records the interface condition.  
In Section~\ref{sec:angular} we perform the angular momentum decomposition and
recall the free radial spectrum.  Section~\ref{sec:green} gives the scalar
Green-function equation and proves monotonicity and uniqueness.  In
Section~\ref{sec:kummer} we obtain an explicit Kummer-function
representation.  Section~\ref{sec:asym} contains the large-$m$ asymptotics
and the main mode-resolved shift formula, including the next algebraic
correction and the uniform repulsive crossover.
Section~\ref{sec:landaupoints} discusses the Landau energies themselves and
their relation to the known results on circular delta interactions.
Section~\ref{sec:numerics} checks the asymptotic formula against the exact
scalar equation, and Section~\ref{sec:discussion} discusses projection
errors, the hard-wall crossover, and boundary response.

\section{The circular $\delta$ interaction}\label{sec:model}

Let
$$
S_a=\{x\in\R^2:|x|=a\},\qquad a>0,
$$
and let $\alpha\in\R$ be constant. We use the magnetic Sobolev space
$$
H_A^1(\R^2)
=\{u\in L^2(\R^2):(-i\nabla-A)u\in L^2(\R^2;\C^2)\}.
$$
The trace on the compact smooth curve $S_a$ is well defined for functions in
$H_A^1(\R^2)$ because this space agrees locally with the usual $H^1$ space.
We define the quadratic form

\begin{equation}\label{eq}
q_{B,\alpha}[u]
=\int_{\R^2}|(-i\nabla-A)u|^2\,d x
+\alpha\int_{S_a}|u|^2\,d s,
\qquad
{\cal D}(q_{B,\alpha})=H_A^1(\R^2).
\end{equation}
Here ${\cal D}(q_{B,\alpha})$ denotes the form domain of $q_{B,\alpha}$.
Also, $[B]=\mathrm{length}^{-2}$ and
$[\alpha]=\mathrm{length}^{-1}$.
For example, the radial potential $V_w(r)=\alpha w^{-1}W((r-a)/w)$,
with real bounded compactly supported $W$ and $\int W(t)\,dt=1$, has this
form limit as $w\downarrow0$ at fixed $\alpha$. The norm resolvent
approximation for such thin magnetic layers is treated in \cite{BEHL2020}.
This fixed-strength thin-layer limit does not assert uniformity in a
simultaneous hard-wall or large-$m$ limit.

The trace term is form bounded with relative bound zero. Hence
$q_{B,\alpha}$ is closed and lower bounded and defines a unique self-adjoint
operator, denoted by $H_{B,\alpha}$. This is the standard quadratic-form
realization of a $\delta$ interaction supported on a curve; see \cite{BEHL2020} and also \cite{AGHH} for general background on singular interactions.

In the symmetric gauge the radial component of $A$ vanishes. Therefore, if
$u$ belongs to the operator domain and has $H^2$ regularity locally up to
the interface on each side, the form identity gives the familiar interface
conditions
\begin{equation}\label{eq:interface}
u(a+0,\theta)=u(a-0,\theta),
\qquad
\partial_r u(a+0,\theta)-\partial_r u(a-0,\theta)
=\alpha u(a,\theta).
\end{equation}
Here $a+0$ and $a-0$ denote the limits from $r>a$ and $r<a$, respectively.
Thus $\alpha>0$ is repulsive and $\alpha<0$ is attractive.

The general theory already implies
$$
\spec_{\ess}(H_{B,\alpha})
=\{\Lambda_n:n\in\N_0\};
$$
see \cite{BEHL2020}. We shall not reprove this general result. Our aim is the
explicit structure generated by circular symmetry.

\section{Angular momentum decomposition}\label{sec:angular}

Write
$$
u(r,\theta)
=\frac{1}{\sqrt{2\pi}}
\sum_{m\in\mathbb Z} f_m(r)e^{im\theta}.
$$
Then
$$
L^2(\R^2)
=\bigoplus_{m\in\mathbb Z}L^2((0,\infty),r\,d r),
$$
and the free operator decomposes into the radial operators
\begin{equation}\label{eq:radial-free}
h_{m,0}
=-\frac{d^2}{d r^2}
-\frac{1}{r}\frac{d}{d r}
+\frac{m^2}{r^2}
+\frac{B^2r^2}{4}-Bm.
\end{equation}
The realization at $r=0$ is inherited from the full-plane quadratic
form. For $m=0$ it excludes the logarithmic solution; no additional point
interaction is placed at the origin. For $m\ne0$ it selects the regular
$r^{|m|}$ behavior. Each channel has compact resolvent because the radial
potential confines at infinity. The $\delta$ interaction preserves the
decomposition and imposes
\begin{equation}\label{eq:radial-jump}
f_m(a+0)=f_m(a-0),
\qquad
f_m'(a+0)-f_m'(a-0)=\alpha f_m(a).
\end{equation}

It is often convenient to use the unitary transformation
$$
v_m(r)=r^{1/2}f_m(r),
$$
from $L^2((0,\infty),r\,d r)$ to $L^2(0,\infty)$. Then
\begin{equation}\label{eq:radial-1d}
\widetilde h_{m,0}
=-\frac{d^2}{d r^2}
+\frac{m^2-1/4}{r^2}
+\frac{B^2r^2}{4}-Bm,
\end{equation}
and the wall becomes the ordinary one-dimensional point interaction
$\alpha\delta(r-a)$. In particular,
$$
v_m'(a+0)-v_m'(a-0)=\alpha v_m(a).
$$

We first record the free radial spectrum.

\begin{prop}\label{prop:radial-spectrum}
For $m\in\mathbb Z$,
$$
\spec(h_{m,0})
=\{\varepsilon_{j,m}:j\in\N_0\},
\qquad
\varepsilon_{j,m}
=B\bigl(2j+|m|-m+1\bigr).
$$
A fixed Landau level $\Lambda_n=B(2n+1)$ occurs in the channel $m$ if and only if
$m\ge -n$.
\end{prop}

\begin{proof}
After the substitution
$$
\rho=\frac{Br^2}{2},
\qquad
f(r)=r^{|m|}e^{-\rho/2}y(\rho),
$$
the radial equation is Kummer's equation. The regular condition at zero and square integrability at
infinity require the first Kummer parameter to be a nonpositive integer,
which gives
$$
\varepsilon_{j,m}
=B\bigl(2j+|m|-m+1\bigr).
$$
The equation $\varepsilon_{j,m}=\Lambda_n$ is equivalent to
$$
j=n-\frac{|m|-m}{2}.
$$
This is a nonnegative integer exactly when $m\ge -n$.
\end{proof}

For the large-mode analysis only $m\ge0$ is needed. In this case
$$
\varepsilon_{j,m}=\Lambda_j
$$
for every $j\in\N_0$. A normalized radial eigenfunction in
$L^2((0,\infty),r\,d r)$ can be written as
\begin{equation}\label{eq:radial-eigenfunction}
f_{j,m}(r)
=N_{j,m}\,r^m e^{-Br^2/4}
L_j^{(m)}\left(\frac{Br^2}{2}\right),
\qquad m\ge0,
\end{equation}
where
\begin{equation}\label{eq:Nnorm}
N_{j,m}^2
=\frac{B^{m+1}j!}{2^m(j+m)!}.
\end{equation}
The corresponding full eigenfunction is
$(2\pi)^{-1/2}f_{j,m}(r)e^{im\theta}$.

\section{Scalar Green-function equation}\label{sec:green}

Let $m\ge0$ and let $E\notin\{\Lambda_j:j\in\N_0\}$. Denote by
$G_m(E;r,s)$ the integral kernel of $(h_{m,0}-E)^{-1}$ in the radial space
$L^2((0,\infty),r\,d r)$, so that
$$
((h_{m,0}-E)^{-1}F)(r)
=\int_0^\infty G_m(E;r,s)F(s)s\,d s.
$$
Define
\begin{equation}\label{eq:gdef}
g_m(E)=aG_m(E;a,a).
\end{equation}
Equivalently, in the one-dimensional representation
\eqref{eq:radial-1d}, $g_m(E)$ is the diagonal resolvent matrix element at
$r=a$.

The spectral expansion gives
\begin{equation}\label{eq:gspectral}
g_m(E)
=\sum_{j=0}^\infty
\frac{d_{j,m}}{\Lambda_j-E},
\qquad
d_{j,m}=a|f_{j,m}(a)|^2\ge0,
\end{equation}
The diagonal series converges locally uniformly away from its active poles.
Indeed, point evaluation is continuous on the radial form domain near $a$,
so $g_m(E_*)<\infty$ for $E_*<B$; comparison with this positive series gives
local convergence, including that of differentiated series. An active pole
means an index $j$ with $d_{j,m}>0$. If $d_{j,m}=0$, that summand is absent
and the singularity at $\Lambda_j$ is removable.

The eigenvalues away from the free radial spectrum are characterized by the following scalar equation.

\begin{prop}\label{prop:scalar-equation}
Let $m\ge0$ and let $E$ be real and different from the free radial eigenvalues.
Then $E$ is an eigenvalue of the perturbed radial operator $h_{m,\alpha}$ if
and only if
\begin{equation}\label{eq:BS}
1+\alpha g_m(E)=0.
\end{equation}
\end{prop}

\begin{proof}
In the one-dimensional representation, an eigenfunction satisfies
$$
(\widetilde h_{m,0}-E)v
=-\alpha\delta_a v(a),
$$
where $\delta_a=\delta(\,\cdot-a)$ is the Dirac delta at $a$.
Applying the free resolvent gives
$$
v=-\alpha(\widetilde h_{m,0}-E)^{-1}\delta_a\,v(a).
$$
Evaluating at $r=a$ yields
$$
v(a)=-\alpha g_m(E)v(a).
$$
If $v(a)=0$, then, since $E$ is assumed not to be a free radial
eigenvalue, the free resolvent equation gives $v=0$. Thus a nonzero
eigenfunction necessarily has $v(a)\ne0$ and satisfies \eqref{eq:BS}. Conversely, if \eqref{eq:BS} holds, the free Green
function with a pole source at $a$ gives an eigenfunction satisfying the jump
condition \eqref{eq:radial-jump}.
\end{proof}

The scalar Green function is monotone on intervals without active poles.

\begin{prop}\label{prop:monotonicity}
For real $E$ on every interval containing no active pole of $g_m$,
$$
\frac{d}{d E}g_m(E)
=\sum_{j=0}^\infty
\frac{d_{j,m}}{(\Lambda_j-E)^2}>0.
$$
Hence the equation \eqref{eq:BS} has at most one solution in each interval
between consecutive active poles.
\end{prop}

\begin{proof}
The derivative formula follows from the spectral representation
\eqref{eq:gspectral}; equivalently it is the positivity of
$(h_{m,0}-E)^{-2}$ in the quadratic-form pairing with the point-evaluation functional. Since the evaluation functional at
$r=a$ is nonzero, the derivative is strictly positive.
\end{proof}

For fixed $n$, the residue of $g_m$ at $\Lambda_n$ is nonzero for all
sufficiently large $m$, because $L_n^{(m)}(x)$ is then positive for fixed
$x>0$. Therefore the neighboring-pole picture is standard for large $m$.

If two consecutive free eigenvalues are active, $g_m$ runs from
$-\infty$ to $+\infty$ in the gap between them. Hence for $\alpha\ne0$
there is exactly one nonfree root in that gap. Below the lowest free
eigenvalue, $g_m$ increases from $0$ to $+\infty$, so there is one root
for $\alpha<0$ and none for $\alpha>0$. The limit at $-\infty$ follows
from the positive series by dominated convergence. These statements concern
the nonfree roots; a free eigenfunction with zero boundary trace must be
handled separately.

This interlacing argument alone does not determine which endpoint a root
approaches as $m$ grows. Existence and uniqueness in a fixed neighborhood
of $\Lambda_n$, together with the direction and rate of convergence, will
be proved in Theorem~\ref{thm:shift}. Form ordering also shows that each
ordered channel eigenvalue is nondecreasing in $\alpha$.

\section{Explicit Kummer representation}\label{sec:kummer}

The scalar equation can also be written directly as a matching condition.
This known representation \cite[Sec.~3.1]{ExnerTater2004} is useful for
numerical calculations and for the large-$m$ estimates. We derive it to fix
the radial normalization and the sign of the coupling.
In that reference, positive coupling denotes attraction, and the displayed
radial magnetic term is $+Bm$. Its formulas therefore correspond to ours
after reversing both the coupling sign and the angular momentum label.

Set
\begin{equation}\label{eq:parameters}
\rho=\frac{Br^2}{2},
\qquad
b_m=|m|+1,
\qquad
\eta_m(E)=\frac{|m|-m+1}{2}-\frac{E}{2B}.
\end{equation}
A solution regular at the origin is
\begin{equation}\label{eq:regular-solution}
\phi_m(r;E)
=r^{|m|}e^{-Br^2/4}
\Mfun\left(\eta_m(E),b_m,\frac{Br^2}{2}\right),
\end{equation}
where $\Mfun$ is Kummer's confluent hypergeometric function. A solution which
is square integrable at infinity is
\begin{equation}\label{eq:decaying-solution}
\psi_m(r;E)
=r^{|m|}e^{-Br^2/4}
\Ufun\left(\eta_m(E),b_m,\frac{Br^2}{2}\right).
\end{equation}
We use the standard notation and identities for $\Mfun$ and $\Ufun$ as in
\cite{NIST,DLMF}.

The radial eigenvalue condition can be written directly as follows.

\begin{prop}\label{prop:matching}
For an energy at which the boundary values in
\eqref{eq:regular-solution} and \eqref{eq:decaying-solution} are both nonzero,
the radial
eigenvalue condition is
\begin{equation}\label{eq:logmatch}
\frac{\psi_m'(a;E)}{\psi_m(a;E)}
-\frac{\phi_m'(a;E)}{\phi_m(a;E)}
=\alpha.
\end{equation}
The following determinant equation is valid also when a boundary value
vanishes, and for every $m\in\mathbb Z$:
\begin{equation}\label{eq:detmatch}
\phi_m(a;E)\psi_m'(a;E)
-\phi_m'(a;E)\psi_m(a;E)
-\alpha\phi_m(a;E)\psi_m(a;E)=0.
\end{equation}
\end{prop}

\begin{proof}
Take a multiple of $\phi_m$ on $(0,a)$ and a multiple of $\psi_m$ on
$(a,\infty)$. Continuity and the derivative jump give a homogeneous two-by-two
linear system for these constants. Its determinant is
\eqref{eq:detmatch}. When both boundary values are nonzero one can divide
to obtain \eqref{eq:logmatch}. If both boundary values vanish, then $E$ is
a free radial eigenvalue. Their nonzero derivatives fix the relative
constants, and the determinant formulation still applies.
\end{proof}

Let
$$
x=\frac{Ba^2}{2}.
$$
For $m\ge0$ one has
$$
\eta_m(E)=\frac12-\frac{E}{2B}.
$$
Using
$$
\frac{\partial}{\partial z}\Mfun(\eta,b,z)
=\frac{\eta}{b}\Mfun(\eta+1,b+1,z),
$$
and
$$
\frac{\partial}{\partial z}\Ufun(\eta,b,z)
=-\eta\Ufun(\eta+1,b+1,z),
$$
equation \eqref{eq:logmatch} becomes an entirely explicit scalar equation.
The common logarithmic derivative of $r^m e^{-Br^2/4}$ cancels, and hence
\begin{eqnarray}
\alpha
&=&Ba\left\{
-\eta_m(E)
\frac{\Ufun(\eta_m(E)+1,m+2,x)}
{\Ufun(\eta_m(E),m+1,x)}\right.\nonumber\\
&&\left.\qquad
-\frac{\eta_m(E)}{m+1}
\frac{\Mfun(\eta_m(E)+1,m+2,x)}
{\Mfun(\eta_m(E),m+1,x)}
\right\}.
\label{eq:kummer-match}
\end{eqnarray}

The Wronskian identity \cite[Eq.~13.2.34]{DLMF}
$$
W_z\{\Mfun(\eta,b,z),\Ufun(\eta,b,z)\}
=-\frac{\Gamma(b)}{\Gamma(\eta)}e^z z^{-b}
$$
leads to a closed expression for the scalar Green function. In that DLMF
identity the first function is the regularized function $M/\Gamma(b)$;
the factor $\Gamma(b)$ above restores our normalization of $M$.

The diagonal Green function has the following explicit form.

\begin{prop}\label{prop:explicit-g}
For $m\ge0$ and $E\notin\{\Lambda_j:j\in\N_0\}$,
\begin{equation}\label{eq:g-kummer}
g_m(E)
=\frac{a e^{-x}x^m}{2m!}
\Gamma(\eta)
\Mfun(\eta,m+1,x)
\Ufun(\eta,m+1,x),
\qquad
\eta=\frac12-\frac{E}{2B}.
\end{equation}
Moreover, the matching equation \eqref{eq:logmatch}, wherever the quotients are defined, and the Green-function
equation \eqref{eq:BS} are identical.
\end{prop}

\begin{proof}
Let $W_r$ denote the Wronskian in the radial variable. The above Kummer
Wronskian gives
$$
rW_r\{\phi_m,\psi_m\}
=-\frac{2^{m+1}m!}{B^m\Gamma(\eta)}.
$$
The radial Green kernel is obtained from the regular and decaying solutions in
the usual Sturm--Liouville form. On the diagonal,
$$
aG_m(E;a,a)
=-\frac{a\phi_m(a;E)\psi_m(a;E)}
{aW_r\{\phi_m,\psi_m\}},
$$
which gives \eqref{eq:g-kummer}. Also,
$$
\frac{\psi_m'}{\psi_m}-\frac{\phi_m'}{\phi_m}
=\frac{W_r\{\phi_m,\psi_m\}}{\phi_m\psi_m}
=-\frac{1}{g_m(E)}.
$$
Therefore \eqref{eq:logmatch} is equivalent to
$g_m(E)=-1/\alpha$, which is \eqref{eq:BS}.
\end{proof}

For $m<0$, the same calculation gives \eqref{eq:g-kummer} with
$x^m/m!$ replaced by $x^{|m|}/|m|!$, the second Kummer parameter
replaced by $|m|+1$, and $\eta$ replaced by $\eta_m(E)$ from
\eqref{eq:parameters}. Its poles are the channel energies
$\varepsilon_{j,m}$, rather than every Landau level.

\section{Large angular momentum and mode-resolved shifts}\label{sec:asym}

We now fix a Landau level $\Lambda_n$ and let $m\to+\infty$. This is the
large-degeneracy direction in the symmetric gauge.

\subsection{Residue at a Landau pole}

Since
$$
\eta(\Lambda_n)=-n,
$$
the gamma factor in \eqref{eq:g-kummer} has a simple pole at
$E=\Lambda_n$. The standard identities \cite[Eq.~13.6.19]{DLMF}
$$
\Mfun(-n,m+1,x)
=\frac{n!m!}{(n+m)!}L_n^{(m)}(x)
$$
and
$$
\Ufun(-n,m+1,x)
=(-1)^n n!L_n^{(m)}(x)
$$
give the pole coefficient explicitly. The residue in the complex variable
$E$ is $-c_{n,m}$, since we use the denominator $\Lambda_n-E$.

The pole coefficient can be interpreted as a boundary weight.

\begin{prop}\label{prop:boundary-weight}
For $m\ge0$,
\begin{equation}\label{eq:cdef}
g_m(E)
=\frac{c_{n,m}}{\Lambda_n-E}+r_{n,m}(E),
\end{equation}
where $r_{n,m}$ is regular near $\Lambda_n$ and
\begin{equation}\label{eq:c-exact}
c_{n,m}
=aB e^{-x}\frac{n!}{(n+m)!}x^m
\bigl[L_n^{(m)}(x)\bigr]^2.
\end{equation}
Equivalently,
$$
c_{n,m}=a|f_{n,m}(a)|^2
=\int_{S_a}|\Psi_{n,m}|^2\,d s,
$$
where $\Psi_{n,m}$ is the normalized free Landau eigenfunction in the
$m$-th channel.
\end{prop}

\begin{proof}
The last identity follows immediately from
\eqref{eq:radial-eigenfunction} and \eqref{eq:Nnorm}. It also follows by inserting
the pole expansion
$$
\Gamma(-n+\varepsilon)
=\frac{(-1)^n}{n!\,\varepsilon}+O(1)
$$
into \eqref{eq:g-kummer} and using
$$
\varepsilon
=\eta+n
=-\frac{E-\Lambda_n}{2B}.
$$
\end{proof}

The quantity $c_{n,m}$ has dimension $\mathrm{length}^{-1}$. It is a
probability density integrated along the circle, not a dimensionless
probability on the zero-area set $S_a$. For fixed $n,m$, its operational
meaning is the thin-annulus relation
\begin{equation}\label{eq:annulus-weight}
\int_{a-w/2<|y|<a+w/2}|\Psi_{n,m}(y)|^2\,d y
=w c_{n,m}+o(w),\qquad w\downarrow0.
\end{equation}

\subsection{Large-$m$ behavior of the residue}

We first record the large-parameter asymptotics of the Laguerre polynomial.

\begin{lem}\label{lem:lag-large-m}
For fixed $n\in\N_0$ and fixed $x>0$,
\begin{equation}\label{eq:lag-large-m}
L_n^{(m)}(x)
=\binom{n+m}{n}
\left(1-\frac{nx}{m+1}+O(m^{-2})\right),
\qquad m\to\infty.
\end{equation}
\end{lem}

\begin{proof}
Use the finite expansion
$$
L_n^{(m)}(x)
=\sum_{j=0}^n\frac{(-x)^j}{j!}
\binom{n+m}{n-j}.
$$
After division by $\binom{n+m}{n}$, the $j=1$ term is
$-nx/(m+1)$, while every term with $j\ge2$ is $O(m^{-2})$ because $n$ is fixed.
\end{proof}

This gives the following superexponential asymptotics of the boundary weight.

\begin{prop}\label{prop:c-asym}
For fixed $n$, $a$, and $B$, as $m\to\infty$,
\begin{equation}\label{eq:c-asym-1}
c_{n,m}
=aB e^{-x}\frac{n!}{(n+m)!}x^m
\binom{n+m}{n}^2
\left(1-\frac{2nx}{m}+O(m^{-2})\right).
\end{equation}
Equivalently,
\begin{equation}\label{eq:c-asym-2}
c_{n,m}
=\frac{aB e^{-x}}{n!}
\frac{x^m m^n}{m!}
\left[1+\frac{\kappa_n(x)}{m}+O(m^{-2})\right],
\end{equation}
where
\begin{equation}\label{eq:kappa}
\kappa_n(x)=\frac{n(n+1)}{2}-2nx.
\end{equation}
In particular, $c_{n,m}$ decreases faster than every exponential in $m$.
\end{prop}

\begin{proof}
Equation~\eqref{eq:lag-large-m} gives
$$
\bigl[L_n^{(m)}(x)\bigr]^2
=\binom{n+m}{n}^2
\left(1-\frac{2nx}{m}+O(m^{-2})\right).
$$
Moreover,
$$
\frac{n!}{(n+m)!}\binom{n+m}{n}^2
=\frac{(n+m)!}{n!(m!)^2}
=\frac{m^n}{n!m!}
\left[1+\frac{n(n+1)}{2m}+O(m^{-2})\right].
$$
Multiplying these two expansions proves \eqref{eq:c-asym-1} and
\eqref{eq:c-asym-2}.
\end{proof}

\subsection{The regular part of the radial Green function}

The residue $c_{n,m}$ is superexponentially small, whereas the regular part of
the diagonal Green function has an algebraic scale.  We now determine its
leading term.  This step is important because the coupling $\alpha$ is fixed
and is not assumed to be small.

The regular part admits the following estimate uniformly near the Landau energy.

\begin{lem}\label{lem:regular-part}
Fix $n\in\N_0$, $a>0$, $B>0$, and $0<\varepsilon<B$.
Uniformly for complex $|E-\Lambda_n|\le\varepsilon$,
\begin{equation}\label{eq:regular-estimate}
r_{n,m}(E)=\frac{a}{2m}+\frac{ax}{2m^2}+O(m^{-3}),
\qquad x=\frac{Ba^2}{2},
\end{equation}
and $r_{n,m}'(E)=O(m^{-3})$. For real $E$ in this interval,
$r_{n,m}'(E)\ge0$.
\end{lem}

\begin{proof}
We first evaluate the resolvent at two real energies below its spectrum.
Write $S_k(x)=\sum_{\ell=0}^k x^\ell/\ell!$. The integral representation
of $U$ \cite[Eq.~13.4.4]{DLMF}, used only at the positive parameters
$\eta=1,2$, gives for $m\ge2$
\begin{eqnarray}
\Gamma(1)U(1,m+1,x)
&=&\Gamma(m)x^{-m}S_{m-1}(x),\label{eq:U-one}\\
\Gamma(2)U(2,m+1,x)
&=&\Gamma(m)x^{-m}
\left[S_{m-1}(x)-\frac{x}{m-1}S_{m-2}(x)\right].
\label{eq:U-two}
\end{eqnarray}
Indeed, the first integral is
$\int_0^\infty e^{-xt}(1+t)^{m-1}\,dt$, whose polynomial expansion gives
\eqref{eq:U-one}. For the second, use
$t(1+t)^{m-2}=(1+t)^{m-1}-(1+t)^{m-2}$.
For $k=1,2$, the defining series gives
\begin{equation}\label{eq:M-reference}
M(k,m+1,x)=1+\frac{kx}{m}+O(m^{-2}).
\end{equation}
To bound the remainder, note that for $\ell\ge2$,
$(m+1)_\ell\ge(m+1)(m+2)(3)_{\ell-2}$. The remaining series, with
numerator $(k)_\ell x^\ell/\ell!$, is summable for fixed $x$.
Here $(q)_\ell$ denotes the Pochhammer symbol.
Also $S_{m-j}(x)=e^x+O(m^{-N})$ for every fixed $j,N$.
Substitution into \eqref{eq:g-kummer} yields
\begin{equation}\label{eq:reference-g}
g_m(-B)=\frac{a}{2m}+\frac{ax}{2m^2}+O(m^{-3}),
\qquad
g_m(-3B)=\frac{a}{2m}+\frac{ax}{2m^2}+O(m^{-3}).
\end{equation}
Consequently, the positive spectral sum satisfies
\begin{equation}\label{eq:positive-sum}
\sum_{j=0}^\infty
\frac{d_{j,m}}{(\Lambda_j+B)(\Lambda_j+3B)}
=\frac{g_m(-B)-g_m(-3B)}{2B}=O(m^{-3}).
\end{equation}

For $j\ne n$ and $|E-\Lambda_n|\le\varepsilon$, the ratios
$(\Lambda_j+3B)/|\Lambda_j-E|$ are bounded uniformly in $j$ and $E$.
The spectral representation therefore gives the exact identity
\begin{eqnarray}
r_{n,m}(E)
&=&g_m(-B)-\frac{c_{n,m}}{\Lambda_n+B}\nonumber\\
&&{}+(E+B)\sum_{j\ne n}
\frac{d_{j,m}}{(\Lambda_j-E)(\Lambda_j+B)}.
\label{eq:regular-reference}
\end{eqnarray}
The sum on the second line is $O(m^{-3})$ by
\eqref{eq:positive-sum}. The removed term is smaller than every power of
$m^{-1}$ by Proposition~\ref{prop:c-asym}. This proves
\eqref{eq:regular-estimate}, including at $E=\Lambda_n$.
Finally,
$$
r_{n,m}'(E)=\sum_{j\ne n}\frac{d_{j,m}}{(\Lambda_j-E)^2}.
$$
The ratios $(\Lambda_j+B)(\Lambda_j+3B)/|\Lambda_j-E|^2$ are uniformly
bounded for $j\ne n$, so \eqref{eq:positive-sum} proves the derivative
bound and the sign assertion.
\end{proof}

We can now state the mode-resolved eigenvalue shift.

\begin{thm}\label{thm:shift}
Fix $n\in\N_0$, $B>0$, $a>0$, and $\alpha\in\R\setminus\{0\}$. For all
sufficiently large $m\ge0$, there is exactly one eigenvalue
$E_{n,m}^{(\alpha)}$ of the $m$-th radial channel in
$(\Lambda_n-\varepsilon,\Lambda_n+\varepsilon)$, where
$0<\varepsilon<B$ is fixed. It satisfies
\begin{equation}\label{eq:shift-c}
E_{n,m}^{(\alpha)}-\Lambda_n
=\alpha c_{n,m}
\left[1-\frac{\alpha a}{2m}+O(m^{-2})\right],
\qquad m\to\infty,
\end{equation}
where $c_{n,m}$ is the exact boundary weight in \eqref{eq:c-exact}. Consequently,
\begin{eqnarray}
E_{n,m}^{(\alpha)}-\Lambda_n
&=&\frac{\alpha aB e^{-x}}{n!}
\frac{x^m m^n}{m!}
\left[1+\frac{\kappa_n(x)-\alpha a/2}{m}
+O(m^{-2})\right],\label{eq:shift-explicit}
\\[-2mm]
&&\hspace{20mm} x=\frac{Ba^2}{2},\qquad
\kappa_n(x)=\frac{n(n+1)}2-2nx.\nonumber
\end{eqnarray}
\end{thm}

\begin{proof}
Put $c=c_{n,m}>0$ for large $m$ and write $E=\Lambda_n+\delta$.
Multiplication of the scalar equation by $\delta$ gives
\begin{equation}\label{eq:Fdelta}
F_m(\delta):=\delta[1+\alpha r_{n,m}(\Lambda_n+\delta)]-\alpha c=0.
\end{equation}
The regular-part lemma gives $|\alpha r_{n,m}|<1/4$ throughout the
fixed neighborhood once $m$ is large. The values of $F_m$ at $\delta=\alpha c/2$ and
$\delta=2\alpha c$ have opposite signs, irrespective of the sign of
$\alpha$.
Both points belong to that neighborhood by Proposition~\ref{prop:c-asym}.
Thus there is a real root with the sign of $\alpha$ and
$|\delta|\le2|\alpha|c$. Moreover,
$$
F_m'(\delta)=1+\alpha r_{n,m}(\Lambda_n+\delta)
+\alpha\delta r_{n,m}'(\Lambda_n+\delta)>0
$$
throughout the neighborhood for large $m$. Hence this root is unique.
The center $\delta=0$ is not an eigenvalue: its free eigenfunction has
nonzero trace, and \eqref{eq:detmatch} at that energy is nonzero.
Any other eigenvalue in the neighborhood would satisfy \eqref{eq:Fdelta}.
The bound on $|\delta|$ also proves convergence to $\Lambda_n$.

At the root we have
\begin{equation}\label{eq:delta-solve}
\delta=\frac{\alpha c_{n,m}}
{1+\alpha r_{n,m}(\Lambda_n+\delta)}.
\end{equation}
Inserting \eqref{eq:regular-estimate} and expanding the denominator proves
\eqref{eq:shift-c}. Together with \eqref{eq:c-asym-2}, this gives
\eqref{eq:shift-explicit}. In fact the same proof gives the stronger form
\begin{equation}\label{eq:shift-second}
\frac{E_{n,m}^{(\alpha)}-\Lambda_n}{\alpha c_{n,m}}
=1-\frac{\alpha a}{2m}
+\frac{(\alpha a/2)^2-\alpha ax/2}{m^2}+O(m^{-3}).
\end{equation}
\end{proof}

The theorem implies one-sided accumulation at a factorial scale.

\begin{cor}\label{cor:factorial}
Under the assumptions of Theorem~\ref{thm:shift},
$$
\operatorname{sgn}\bigl(E_{n,m}^{(\alpha)}-\Lambda_n\bigr)
=\operatorname{sgn}(\alpha)
$$
for all sufficiently large $m$, and
\begin{equation}\label{eq:root-scale}
\lim_{m\to\infty}
\left(m!\left|E_{n,m}^{(\alpha)}-\Lambda_n\right|\right)^{1/m}
=\frac{Ba^2}{2}.
\end{equation}
Moreover, if the exact boundary weight is used for normalization, then
\begin{equation}\label{eq:next-order-limit}
\lim_{m\to\infty}
m\left(
\frac{E_{n,m}^{(\alpha)}-\Lambda_n}{\alpha c_{n,m}}-1
\right)
=-\frac{\alpha a}{2}.
\end{equation}
\end{cor}

\begin{proof}
The sign and \eqref{eq:next-order-limit} follow from \eqref{eq:shift-c}.
Formula \eqref{eq:root-scale} follows from \eqref{eq:shift-explicit}, since all
factors other than $x^m/m!$ have $m$-th root tending to one.
\end{proof}

We briefly relate this scale to the known capacity estimates.

\begin{rem}\label{rem:capacity}
For a circle of radius $a$, the logarithmic capacity is $a$. The upper
and lower estimates in \cite[Theorem~6.4 and Proposition~6.6]{BEHL2020},
applied to the circle and to
closed subarcs whose capacities tend to $a$, give the root scale $Ba^2/2$. Thus the circle scale is
already contained in those results; the angular labels and algebraic
corrections in \eqref{eq:shift-explicit} give more detailed information.
\end{rem}

\subsection{Uniform repulsive crossover and the hard wall}
\label{sec:crossover}

The fixed-$\alpha$ expansion cannot be used by sending $\alpha$ to
infinity term by term. Keeping the denominator in \eqref{eq:delta-solve}
resolves this issue and connects the penetrable model with the exterior
Dirichlet setting of \cite{PushnitskiRozenblum2007}.

The repulsive crossover is uniform up to the hard-wall limit.

\begin{thm}\label{thm:crossover}
Fix $n$, $a$, $B$, and $0<\varepsilon<B$. For all sufficiently large $m$,
independently of $\alpha>0$, there is exactly one eigenvalue in
$(\Lambda_n,\Lambda_n+\varepsilon)$. Set
$$
\beta=\frac{\alpha a}{2m},\qquad
\delta_{n,m}^{(\alpha)}=E_{n,m}^{(\alpha)}-\Lambda_n.
$$
Then
\begin{equation}\label{eq:crossover}
\delta_{n,m}^{(\alpha)}
=\frac{2m}{a}c_{n,m}\frac{\beta}{1+\beta}
\left[1-\frac{x}{m}\frac{\beta}{1+\beta}+O(m^{-2})\right],
\end{equation}
where the relative remainder is uniform for $0<\beta<\infty$.
At $\alpha=0$ the shift is zero. In the limit $\alpha\to+\infty$,
the eigenvalue converges to the exterior radial Dirichlet eigenvalue
$E_{n,m}^{D}$ near $\Lambda_n$, and
\begin{equation}\label{eq:dirichlet-shift}
E_{n,m}^{D}-\Lambda_n
=\frac{2m}{a}c_{n,m}\left[1-\frac{x}{m}+O(m^{-2})\right].
\end{equation}
\end{thm}

\begin{proof}
For real $|\delta|\le\varepsilon$, the regular-part lemma gives
$r_{n,m}(\Lambda_n+\delta)\ge a/(4m)>0$ once $m$ is sufficiently large.
At $\delta=0$ the function in \eqref{eq:Fdelta} is negative, whereas at
$d_m=4mc_{n,m}/a$ it is positive, since $d_m r_{n,m}\ge c_{n,m}$.
For large $m$, $d_m<\varepsilon$. On the positive interval,
$F_m'=1+\alpha r_{n,m}+\alpha\delta r_{n,m}'>0$, so the root is unique.
On the negative interval $F_m<0$, and there is no second root near the pole.
In particular, the root satisfies $0<\delta\le d_m$ uniformly in
$\alpha>0$.

The exact denominator in \eqref{eq:delta-solve} is
$$
1+\beta+\frac{\beta x}{m}+O(\beta m^{-2}).
$$
After division by $1+\beta$, the remainder is bounded independently of
$\beta$, because $0\le\beta/(1+\beta)\le1$. Inverting gives
\eqref{eq:crossover}.

As $\alpha\to+\infty$, monotone form convergence imposes the condition
$f(a)=0$ and decouples the interior and exterior radial problems.
Compactness of the channel resolvent and the variational principle give
convergence of each ordered eigenvalue to the corresponding eigenvalue
of this direct sum. For $m>x$, the interior Dirichlet spectrum is bounded
below by
$$
\min_{0<r\le a}\left(\frac{m}{r}-\frac{Br}{2}\right)^2
=\left(\frac{m}{a}-\frac{Ba}{2}\right)^2,
$$
which tends to infinity. The eigenvalue near fixed $\Lambda_n$ must
therefore belong to the exterior problem. Equivalently, taking
$\alpha\to\infty$ in \eqref{eq:delta-solve} gives
$\delta r_{n,m}(\Lambda_n+\delta)=c_{n,m}$, or $g_m(E)=0$.
Since the regular interior solution has no zero there, this is precisely
$\psi_m(a;E)=0$. Letting $\beta\to\infty$ in the uniform estimate proves
\eqref{eq:dirichlet-shift}.
\end{proof}

\section{The Landau energies: nodal states and level crossings}
\label{sec:landaupoints}

A free Landau eigenfunction whose trace on $S_a$ is zero remains an
eigenfunction for every $\alpha$. In a channel $m\ge-n$, define
$$
j=n-\frac{|m|-m}{2}.
$$
The necessary and sufficient condition for this nodal survival in that
channel is
\begin{equation}\label{eq:lag-zero}
L_j^{(|m|)}(x)=0,\qquad x=Ba^2/2.
\end{equation}
Indeed, at $E=\Lambda_n$ the regular and decaying solutions are
proportional to the same free radial eigenfunction. If its boundary value
is nonzero, continuity forces a globally proportional solution, whose
zero derivative jump is incompatible with $\alpha\ne0$. If the value
is zero, equality of the nonzero derivatives fixes the relative constants
and the free eigenfunction satisfies the interface condition. For $m\ge0$,
\eqref{eq:lag-zero} is $L_n^{(m)}(x)=0$; for $m=-k$, $1\le k\le n$,
it is $L_{n-k}^{(k)}(x)=0$.

The known circle theorem concerns the kernel of the Landau-compressed
interaction: for $n\ge1$ its dimension is at most $n$, and the radii for
which it is nonzero form an infinite discrete set
\cite[Theorem~3.9]{BHLR2022}. This kernel describes the states just
identified, rather than every possible eigenfunction of the perturbed
operator at $\Lambda_n$ for arbitrary coupling.

There is another possibility for an attractive wall. If $m<-n$, then
$\Lambda_n$ lies strictly below the free spectrum in that channel.
Define $g_m$ there by the same radial resolvent construction, now with
poles $\varepsilon_{j,m}$. Its value at $\Lambda_n$ is strictly positive,
and the scalar equation gives the coupling
\begin{equation}\label{eq:crossing-coupling}
\alpha_{n,m}^{*}=-\frac{1}{g_m(\Lambda_n)}<0.
\end{equation}
At this value a perturbed channel eigenvalue crosses $\Lambda_n$.
It is not a surviving free Landau state. These are the only additional
possibilities for the constant circular coupling, since the channels
$m\ge-n$ were treated above.

In particular, $L_0^{(m)}=1$ rules out nodal survival at $\Lambda_0=B$.
A repulsive circle therefore has no eigenfunction at $B$, but an
attractive one can have a crossing from a negative angular momentum
channel. For a concrete example, in channel $m=-1$ the Kummer parameter
at $E=B$ is $1$ and
$$
g_{-1}(B)=\frac{a(1-e^{-x})}{2x},\qquad
\alpha_{0,-1}^{*}=-\frac{2x}{a(1-e^{-x})}.
$$
This distinction agrees with \cite[Corollary~3.7]{BHLR2022}, which rules
out the lowest Landau eigenvalue for repulsive coupling but requires a
smallness condition to rule it out for attractive coupling. Regardless of
whether an eigenfunction exists exactly at $\Lambda_n$, that point remains
in the essential spectrum. The fixed-radius, large-$m$ analysis is
unaffected: $L_n^{(m)}(x)>0$ for all sufficiently large $m$.

\section{Numerical check of the large-$m$ expansion}\label{sec:numerics}

We now compare the exact scalar equation with the next-order asymptotic formula.
This calculation is not needed for the proof of Theorem~\ref{thm:shift}; its
purpose is to display the mode dependence on the scale on which the first
algebraic correction is visible.

We take
$$
B=1,\qquad a=1.5,\qquad n=0,
$$
and solve \eqref{eq:BS} with the exact Kummer representation
\eqref{eq:g-kummer} for $\alpha=1$ and $\alpha=-1$.  Define
\begin{equation}\label{eq:Rdef}
R_m^{(\alpha)}
=\frac{E_{0,m}^{(\alpha)}-\Lambda_0}{\alpha c_{0,m}}.
\end{equation}
For $n=0$, Corollary~\ref{cor:factorial} predicts
$$
m\bigl(R_m^{(\alpha)}-1\bigr)
\longrightarrow -\frac{\alpha a}{2}.
$$
Thus the two limits are $-0.75$ and $0.75$ for $\alpha=1$ and $\alpha=-1$,
respectively.

\begin{figure}[t]
\centering
\includegraphics[width=0.95\textwidth]{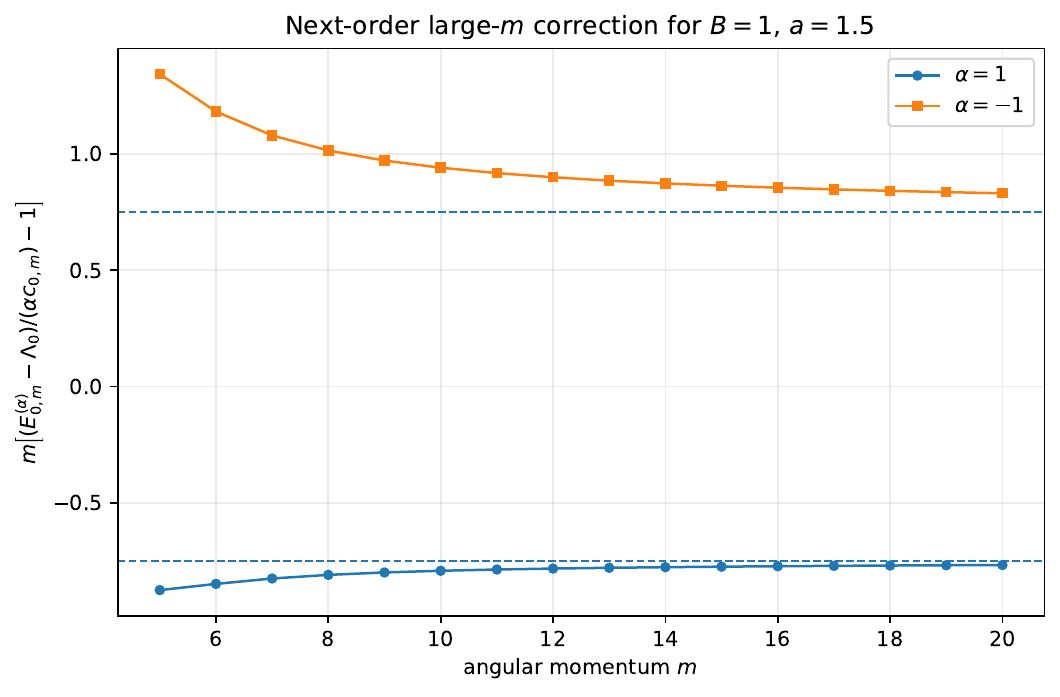}
\caption{Numerical check of the next-order asymptotic formula for $B=1$,
$a=1.5$, and $n=0$.  The plotted quantity is
$m[(E_{0,m}^{(\alpha)}-\Lambda_0)/(\alpha c_{0,m})-1]$.
The dashed horizontal lines are the limits $-\alpha a/2$, namely $-0.75$ for
$\alpha=1$ and $0.75$ for $\alpha=-1$.}
\label{fig:next-order}
\end{figure}

Figure~\ref{fig:next-order} shows the numerical roots for $5\le m\le20$.
The values approach the predicted limits from below for $\alpha=1$ and
from above for $\alpha=-1$. The next term in \eqref{eq:shift-second}
predicts, respectively,
$$
m(R_m^{(1)}-1)=-0.75-\frac{0.28125}{m}+O(m^{-2}),
\qquad
m(R_m^{(-1)}-1)=0.75+\frac{1.40625}{m}+O(m^{-2}).
$$
This explains the unequal rates of approach in the two curves.

Because the shift rapidly becomes smaller than machine precision relative
to $\Lambda_n$, we solve directly for $t=(E-\Lambda_n)/c_{n,m}$.
The pole can be removed analytically using $z=\delta/(2B)$ and
\begin{equation}\label{eq:stable-gamma}
\delta\Gamma(-n-z)
=\frac{-2B\Gamma(1-z)}{\prod_{k=1}^n(-k-z)},
\qquad \delta=E-\Lambda_n,
\end{equation}
with the empty product equal to one and the value at $\delta=0$
understood by continuation. Let
$K_{n,m}(t)=\delta g_m(\Lambda_n+\delta)/c_{n,m}$, evaluated by
\eqref{eq:g-kummer} and \eqref{eq:stable-gamma}; then $K_{n,m}(0)=-1$.
We solve $t/\alpha+K_{n,m}(t)=0$ for finite $\alpha$, and
$K_{n,m}(t)=0$ for the exterior Dirichlet limit. This avoids subtracting
two nearly equal energies. Calculations use \texttt{mpmath}~1.4.1 with
130 decimal digits. The normalized residual is below $10^{-65}$, and the
roots are checked independently in the logarithmic matching equation for
finite coupling and in $U(\eta,m+1,x)=0$ for the Dirichlet limit.
Repeating selected calculations at 170 digits, for both coupling signs
and for $n=2,m=40$ including the Dirichlet limit, gives agreement to more
than 60 relative decimal digits. The extra precision is needed also to
resolve the small displacement of $\eta$ from $-n$ when $n>0$.
If direct evaluation of $U$ fails near a zero, the script uses its
recurrence in the first parameter \cite[Eq.~13.3.7]{DLMF}, starting
from positive parameters and retaining extra guard digits.

We also test the uniform repulsive formula. Table~\ref{tab:crossover}
compares the normalized shift
$D_{n,m}=a(E_{n,m}^{(\alpha)}-\Lambda_n)/(2mc_{n,m})$
with its leading and corrected approximations. The values $\beta=0.1$,
$1$, and $10$ span the overlap regime and the approach to saturation;
$\beta=\infty$ denotes the exterior Dirichlet root. For $n=0$, doubling
$m$ from 20 to 40 reduces the absolute error of the corrected formula
by factors between 4.2 and 4.4 in the displayed cases. The $n=2$ rows
also approach the same crossover law, with a larger remainder. The
constants in the large-$m$ estimates are allowed to depend on $n$.

\begin{table}[t]
\centering
\caption{Repulsive crossover for $B=1$, $a=1.5$ ($x=1.125$).
The leading approximation is $p=\beta/(1+\beta)$ and the corrected
one is $p(1-xp/m)$. The numerical column is obtained from the exact
Kummer equation.}\label{tab:crossover}
\small
\begin{tabular}{rrcrrr}
\toprule
$n$ & $m$ & $\beta$ & Numerical $D_{n,m}$ & Leading & Corrected \\
\midrule
0 & 20 & 0.1 & 0.09038997 & 0.09090909 & 0.09044421 \\
0 & 20 & 1 & 0.48468993 & 0.50000000 & 0.48593750 \\
0 & 20 & 10 & 0.85971608 & 0.90909091 & 0.86260331 \\
0 & 20 & $\infty$ & 0.94057919 & 1.00000000 & 0.94375000 \\
0 & 40 & 0.1 & 0.09066408 & 0.09090909 & 0.09067665 \\
0 & 40 & 1 & 0.49267710 & 0.50000000 & 0.49296875 \\
0 & 40 & 10 & 0.88516964 & 0.90909091 & 0.88584711 \\
0 & 40 & $\infty$ & 0.97113122 & 1.00000000 & 0.97187500 \\
2 & 40 & 0.1 & 0.09063871 & 0.09090909 & 0.09067665 \\
2 & 40 & 1 & 0.49192889 & 0.50000000 & 0.49296875 \\
2 & 40 & 10 & 0.88275734 & 0.90909091 & 0.88584711 \\
2 & 40 & $\infty$ & 0.96822841 & 1.00000000 & 0.97187500 \\
\bottomrule
\end{tabular}
\end{table}

\section{Physical interpretation}\label{sec:discussion}

\subsection{Boundary overlap and the orbit location}

The magnetic length is $\ell_B=B^{-1/2}$ in our units, and
$x=a^2/(2\ell_B^2)$ is the flux through the circle in units of the flux
quantum $2\pi$. In the lowest Landau level the normalized radial
probability, expressed in $\rho=Br^2/2$, is exactly
\begin{equation}\label{eq:gamma-probability}
|f_{0,m}(r)|^2r\,dr
=e^{-\rho}\frac{\rho^m}{m!}\,d\rho.
\end{equation}
Thus $c_{0,m}=aB e^{-x}x^m/m!$ samples the tail of this distribution
at the wall. For a fixed higher index $n$, the Laguerre recurrence and
orthogonality similarly give
\begin{equation}\label{eq:mean-radius}
\langle r^2\rangle_{n,m}=\frac{2}{B}(m+2n+1).
\end{equation}
The same recurrence gives
$\operatorname{Var}(\rho)=(2n+1)m+2n^2+2n+1$.
Thus the mass lies on a radial scale $r\sim\sqrt{2m}\ell_B$, with width
of order $\ell_B$ for fixed $n$. Increasing $m$ at fixed $n$ moves the
orbit distribution outward while leaving the free cyclotron energy
unchanged. This is a change in the guiding-center degree of freedom,
not an increase of the cyclotron excitation index.

For a wall of fixed radius, the limit $m\to\infty$ therefore describes
states far outside the wall. Their shifts are factorially small because
only a tail of the wavefunction reaches it. States near the wall satisfy
roughly $m\sim x$ in the lowest Landau level; the fixed-$x$ expansion is
not uniform in that regime, although the exact scalar equation remains
available. The present asymptotics consequently quantify the remote tail
of the wall-induced dispersion, rather than the entire edge-state spectrum.

\subsection{Projection onto one Landau level and radial mixing}

Let $P_n$ be the free Landau projection and let $T_n$ denote the operator
on $P_nL^2(\R^2)$ associated with the boundary form
$\int_{S_a}|u|^2\,ds$. On the sector $m\ge0$, rotational symmetry makes
$T_n$ diagonal in the angular basis, with eigenvalues $c_{n,m}$. The full
Landau subspace also contains the finitely many channels $-n\le m<0$.
They are omitted here because the following discussion concerns the
large-$m$ behavior. For $m\ge0$, restriction to this Landau subspace gives
the energy
\begin{equation}\label{eq:projected-energy}
E_{n,m}^{\mathrm{proj}}=\Lambda_n+\alpha c_{n,m}.
\end{equation}
This is also first-order perturbation theory in $\alpha$, but
Theorem~\ref{thm:shift} shows its validity at fixed, possibly non-small
$\alpha$ as $m$ grows. The error relative to the projected shift is
\begin{equation}\label{eq:projection-error}
\frac{E_{n,m}^{(\alpha)}-E_{n,m}^{\mathrm{proj}}}
{\alpha c_{n,m}}
=-\frac{\alpha a}{2m}+O(m^{-2}).
\end{equation}

The complementary radial levels produce the regular part
$$
r_{n,m}(\Lambda_n)=\sum_{j\ne n}\frac{d_{j,m}}{\Lambda_j-\Lambda_n}.
$$
The exact denominator in \eqref{eq:delta-solve} incorporates these levels
before any expansion in the coupling. At weak coupling its second-order
contribution is
$-\alpha^2c_{n,m}r_{n,m}(\Lambda_n)$, in agreement with the usual sum
$\sum_{j\ne n}|\langle j,m|\alpha\delta(r-a)|n,m\rangle|^2/
(\Lambda_n-\Lambda_j)$. For large $m$ the regular part is positive;
the correction reduces the repulsive shift and increases the magnitude
of the attractive shift. Although any fixed collection of other Landau
levels has very small boundary weights, their full sum contributes on an
algebraic scale. Keeping that sum is essential for a relative error estimate.

The accuracy of the energy shift must be distinguished from the weight of
the state outside the chosen Landau level. The same resolvent gives an
exact expression for this weight.

The following proposition quantifies the Landau-subspace weight.

\begin{prop}\label{prop:landau-weight}
Let $\Psi_{n,m}^{(\alpha)}$ be the normalized channel eigenfunction
in Theorem~\ref{thm:shift} or Theorem~\ref{thm:crossover}, and set
$\delta=E_{n,m}^{(\alpha)}-\Lambda_n$. Then
\begin{equation}\label{eq:landau-weight}
\|(1-P_n)\Psi_{n,m}^{(\alpha)}\|^2
=\frac{\delta^2 r_{n,m}'(E_{n,m}^{(\alpha)})}
{c_{n,m}+\delta^2 r_{n,m}'(E_{n,m}^{(\alpha)})}.
\end{equation}
For fixed $\alpha\ne0$ this weight is $O(a c_{n,m}m^{-3})$.
Uniformly for all $\alpha>0$ it is $O(a c_{n,m}m^{-1})$, with
$n$, $a$, and $B$ fixed. Both bounds tend to zero factorially.
\end{prop}

\begin{proof}
In the one-dimensional representation the eigenfunction is proportional
to $(\widetilde h_{m,0}-E)^{-1}\delta_a$. Its squared norm before
normalization is
$$
g_m'(E)=\frac{c_{n,m}}{\delta^2}+r_{n,m}'(E).
$$
The squared coefficient of the free $n$-th radial eigenfunction is
$c_{n,m}/\delta^2$. Rotational symmetry identifies this coefficient
with the full Landau projection in the given channel, so subtraction
of its normalized value from one proves \eqref{eq:landau-weight}.
For fixed $\alpha$, use $\delta=O(c_{n,m})$ and
$r_{n,m}'=O(m^{-3})$. For all $\alpha>0$, the proof of
Theorem~\ref{thm:crossover} gives
$0<\delta\le4mc_{n,m}/a$ with a constant independent of $\alpha$.
Together with the same derivative estimate, this proves the uniform
bound. The factors of $a$ express the estimates through the dimensionless
weight $a c_{n,m}$. The implicit constants may depend on the fixed
parameters $n$, $a$, and $B$, and the first bound may also depend on the
fixed value of $\alpha$.
\end{proof}

Consequently, an order-one relative change of the projected energy shift
does not require an order-one admixture of other Landau levels. A small
change in the state norm can substantially change its already very small
boundary trace. This distinction is relevant when projection is used
to predict the dispersion of remote modes or their response to a thin
barrier: norm accuracy alone does not control the relative error of
these small quantities.

\subsection{When a penetrable wall behaves as a hard wall}

The dimensionless parameter controlling this comparison is
$\beta=\alpha a/(2m)$. For $\beta\ll1$, the projected shift is accurate.
When $\beta$ is of order one, \eqref{eq:crossover} reduces the projected
shift by a factor approaching $1/(1+\beta)$, even though the absolute
shift still tends to zero factorially. For $\beta\gg1$, the shift
saturates at the exterior Dirichlet value. In particular,
\begin{equation}\label{eq:wall-ratio}
\frac{E_{n,m}^{(\alpha)}-\Lambda_n}{E_{n,m}^{D}-\Lambda_n}
=\frac{\beta}{1+\beta}\,[1+O(m^{-1})]
\end{equation}
uniformly for $\alpha>0$.

This scale also follows from the radial equation. At the wall the leading
positive potential is $m^2/a^2$, so the local inverse decay length is
$m/a$. A one-dimensional constant-potential resolvent has diagonal value
$1/(2\sqrt{m^2/a^2-E})\sim a/(2m)$. This local estimate explains the
coefficient in the rigorous regular-part lemma. The jump strength must
be comparable with twice this inverse decay length to cause an order-one
relative change. For a fixed finite barrier, increasing $m$ restores the
overlap approximation; reaching the hard-wall value uniformly over such
modes requires a correspondingly increasing strength. The uniform
crossover theorem applies to repulsion. It does not extend through
$1+\alpha a/(2m)\simeq0$ for attraction, where the denominator can be
small and a different analysis would be needed.

\subsection{Boundary response of the perturbed state}

For a normalized simple channel eigenfunction $v_{n,m}^{(\alpha)}$ in
$L^2(0,\infty)$, differentiation of the form gives
\begin{equation}\label{eq:HF}
\frac{\partial E_{n,m}^{(\alpha)}}{\partial\alpha}
=|v_{n,m}^{(\alpha)}(a)|^2
=\int_{S_a}|\Psi_{n,m}^{(\alpha)}|^2\,ds.
\end{equation}
For a nonfree eigenvalue with $\alpha\ne0$, implicit differentiation of
the scalar equation gives
$$
\partial_\alpha E=\frac{1}{\alpha^2g_m'(E)}>0.
$$
More usefully, differentiating
\eqref{eq:delta-solve} implicitly gives
\begin{equation}\label{eq:boundary-response}
\frac{\partial E_{n,m}^{(\alpha)}}{\partial\alpha}
=\frac{c_{n,m}}
{[1+\alpha r_{n,m}(E)]^2+\alpha^2c_{n,m}r_{n,m}'(E)}.
\end{equation}
By Lemma~\ref{lem:regular-part}, for fixed $\alpha$ this becomes
$$
|v_{n,m}^{(\alpha)}(a)|^2
=c_{n,m}\left[1-\frac{\alpha a}{m}+O(m^{-2})\right].
$$
For all $\alpha>0$, the same bounds and
$c_{n,m}r_{n,m}'/r_{n,m}^2=O(c_{n,m}/m)$ yield the uniform version
\begin{equation}\label{eq:response-crossover}
|v_{n,m}^{(\alpha)}(a)|^2
=\frac{c_{n,m}}{(1+\beta)^2}
\left[1-\frac{2x}{m}\frac{\beta}{1+\beta}+O(m^{-2})\right].
\end{equation}
The energy response thus measures the actual boundary density and
shows its suppression as the repulsive wall becomes impenetrable.
For fixed $m$ and finite $\alpha$, a narrow annulus of width $w$ has
probability $w\,\partial_\alpha E_{n,m}^{(\alpha)}+o(w)$, by the same
trace interpretation as in \eqref{eq:annulus-weight}.
This conclusion follows from the stationary single-particle problem;
no assumption about a transport measurement or finite spectral linewidth
is made.

\section{Conclusion}\label{sec:conclusion}

Starting from the known scalar matching equation for a magnetic soft ring,
we have obtained a uniform estimate for the radial resolvent after removal
of one Landau pole. The factorial boundary weight and the algebraic
regular part determine the individual eigenvalue shifts, with an explicit
relative correction to the projected Landau energy. The proof establishes
local existence and uniqueness, and the regular-part estimate remains
valid at the Landau energy itself.

For repulsive coupling the denominator can be retained uniformly up to
the hard-wall limit. The parameter $\alpha a/(2m)$ separates the overlap
regime from saturation at the exterior Dirichlet shift. The same estimate
determines the response to wall strength and hence the probability density
on the boundary. At the same time, the eigenfunction remains close in norm
to the original Landau subspace, even where projection has an order-one
relative error in the energy shift. These results supplement the established cluster and
capacity theory with a quantitative comparison between penetrable and
impenetrable circular barriers. They apply at fixed radius and fixed
Landau index; modes centered on the wall and attraction tuned near a
small denominator remain separate regimes.

\section*{Data availability}
All data that support the findings of this study are included within the
article and its supplementary information files. The numerical values and a
Python script using \texttt{mpmath} to reproduce them are supplied as
supplementary material
(\texttt{check\_modes.py}, \texttt{fixed\_strength.csv},
\texttt{crossover.csv}, and \texttt{precision\_checks.csv}). They are generated
from \eqref{eq:g-kummer} using the pole removal in
\eqref{eq:stable-gamma}; no external data sets are used.

\section*{Acknowledgements}
The author received no specific funding for this work and declares no
conflicts of interest.


\begin{thebibliography}{99}

\bibitem{LandauLifshitz}
L. D. Landau and E. M. Lifshitz,
\textit{Quantum Mechanics: Non-Relativistic Theory},
3rd ed., Pergamon Press, Oxford, 1977.

\bibitem{RaikovWarzel2002}
G. D. Raikov and S. Warzel,
Quasi-classical versus non-classical spectral asymptotics for magnetic
Schr\"odinger operators with decreasing electric potentials,
\textit{Rev. Math. Phys.} {\bfseries 14} (2002), 1051--1072.
\url{https://doi.org/10.1142/S0129055X02001491}.

\bibitem{MelgaardRozenblum2003}
M. Melgaard and G. Rozenblum,
Eigenvalue asymptotics for weakly perturbed Dirac and Schr\"odinger operators
with constant magnetic fields of full rank,
\textit{Commun. Partial Differential Equations} {\bfseries 28} (2003), 697--736.
\url{https://doi.org/10.1081/PDE-120020493}.

\bibitem{PushnitskiRaikovVillegas2013}
A. Pushnitski, G. Raikov, and C. Villegas-Blas,
Asymptotic density of eigenvalue clusters for the perturbed Landau Hamiltonian,
\textit{Commun. Math. Phys.} {\bfseries 320} (2013), 425--453.
\url{https://doi.org/10.1007/s00220-012-1643-4}.

\bibitem{ExnerTater2004}
P. Exner and M. Tater,
Spectra of soft ring graphs,
\textit{Waves Random Media} {\bfseries 14} (2004), S47--S60.
\url{https://doi.org/10.1088/0959-7174/14/1/010}.

\bibitem{BEHL2020}
J. Behrndt, P. Exner, M. Holzmann, and V. Lotoreichik,
The Landau Hamiltonian with $\delta$-potentials supported on curves,
\textit{Rev. Math. Phys.} {\bfseries 32} (2020), no.~4, 2050010.
\url{https://doi.org/10.1142/S0129055X20500105}.

\bibitem{BHLR2022}
J. Behrndt, M. Holzmann, V. Lotoreichik, and G. Raikov,
The fate of Landau levels under $\delta$-interactions,
\textit{J. Spectr. Theory} {\bfseries 12} (2022), no.~3, 1203--1234.
\url{https://doi.org/10.4171/JST/422}.

\bibitem{PushnitskiRozenblum2007}
A. Pushnitski and G. Rozenblum,
Eigenvalue clusters of the Landau Hamiltonian in the exterior of a compact domain,
\textit{Doc. Math.} {\bfseries 12} (2007), 569--586.
\url{https://doi.org/10.4171/DM/235}.

\bibitem{AGHH}
S. Albeverio, F. Gesztesy, R. H\o egh-Krohn, and H. Holden,
\textit{Solvable Models in Quantum Mechanics}, 2nd ed.,
AMS Chelsea Publishing, Providence, RI, 2005.

\bibitem{NIST}
F. W. J. Olver, D. W. Lozier, R. F. Boisvert, and C. W. Clark, eds.,
\textit{NIST Handbook of Mathematical Functions},
Cambridge University Press, New York, 2010.

\bibitem{DLMF}
NIST Digital Library of Mathematical Functions,
Chapter 13, Confluent Hypergeometric Functions, and Chapter 18,
Orthogonal Polynomials.
\url{https://dlmf.nist.gov/13} and \url{https://dlmf.nist.gov/18}
(accessed 6 September 2026).

\end{thebibliography}
\end{document}